# Macromagnetic simulation for reservoir computing utilizing spin-dynamics in magnetic tunnel junctions

Taishi Furuta[1], Keisuke Fujii[2,3‡], Kohei Nakajima[4,3], Sumito Tsunegi[5], Hitoshi Kubota[5], Yoshishige Suzuki[1,6], and Shinji Miwa[1,6,7†]

[1]*Graduate School of Engineering Science, Osaka University, Toyonaka, Osaka 560-8531, Japan*

[2]*Graduate School of Faculty of Science, Kyoto University, Sakyo, Kyoto 606-8502, Japan*

[3]*PRESTO, Japan Science and Technology Agency (JST), Kawaguchi, Saitama 332-0012, Japan*

[4]*Graduate School of Information Science and Technology, The University of Tokyo, Bunkyo, Tokyo 113-8656, Japan*

[5]*National Institute of Advanced Industrial Science and Technology (AIST), Spintronics Research Center, Tsukuba, Ibaraki 305-8568, Japan*

[6]*Center for Spintronics Research Network, Osaka University, Toyonaka, Osaka 560-8531, Japan*

[7]*The Institute for Solid State Physics, The University of Tokyo, Kashiwa, Chiba 277-8581, Japan*

[†]miwa@issp.u-tokyo.ac.jp

[‡]fujii.keisuke.2s@kyoto-u.ac.jp

**ABSTRACT-** The figures-of-merit for reservoir computing (RC), using spintronics devices called magnetic tunnel junctions (MTJs), are evaluated. RC is a type of recurrent neural network. The input information is stored in certain parts of the reservoir, and computation can be performed by optimizing a linear transform matrix for the output. While all the network characteristics should be controlled in a general recurrent neural network, such optimization is not necessary for RC. The reservoir only has to possess a non-linear response with memory effect. In this paper, macromagnetic simulation is conducted for the spin-dynamics in MTJs, for reservoir computing. It is determined that the MTJ-system possesses the memory effect and non-linearity required for RC. With RC using 5-7 MTJs, high performance can be obtained, similar to an echo-state network with 20-30 nodes, even if there are no magnetic and/or electrical interactions between the magnetizations.

## I. INTRODUCTION

The magnetization direction of ferromagnetic metallic film is determined by the magnetic anisotropy energy, which causes non-volatility. This property can be used for magnetic random-access memory devices [1]. In magnetic tunnel junction (MTJ) devices consisting of ferromagnetic and dielectric thin films, the magnetization direction in the ferromagnet can be detected by the change in device resistance originating from the tunneling magnetoresistance (TMR) effect [2-5]. Moreover, the magnetization direction can be electrically controlled by the spin-torque [6-9]. Therefore, MTJ devices are suitable for constructing non-volatile high-density memory devices. In addition to such *long-term* memory effect, the magnetization precessional dynamics appear to possess *short-term* memory effect with non-linear behavior. Such additional magnetization dynamics properties may be suitable for computation using MTJ devices.

The recurrent neural network (RNN) [10, 11] is a machine learning method. It is a mathematical model, which emulates the nerve system in human brain. The RNN concept is depicted in Fig. 1(a). The model consists of three layers, input, middle (node), and output. In the RNN, the information of the middle layer recursively propagates in itself. The middle-layer state is determined by the present input and past middle-layer state, i.e., the middle layer in the RNN possesses memory effect. All the weight matrices for the input ($W_{in}$), middle ($W$) and output ($W_{out}$) should be precisely trained to obtain the desired output. However, when the middle layer has sufficient memory effect and non-linearity, it is feasible to perform computation by optimizing only the output matrix ($W_{out}$). This type of simple RNN is called reservoir computing (RC) [12-14]. In RC, as system training is simple, it is easy to construct large-scale systems. The RC concept is depicted in Fig. 1(b). It has been reported that RC can be implemented in real physical systems, such as atomic switches [15-18], optoelectronic architecture [19-21], and the mechanical bodies of soft and compliant robots [22-24]. While it is possible to perform RC with such classical systems, RC using quantum dynamics can show higher figures-of-merit [25]. Recently, voice recognition by RC using an MTJ [26] was reported; however, the figures-of-merit for RC using MTJs are not quantitatively understood. In this paper, we report the quantitative analysis of the figures-of-merit for RC [27,28] using MTJ devices. We employ macromagnetic simulation for the study.

## II. METHODS

### A. Reservoir computing

The RC method is as follows. A Boolean-type input $s_{\text{in}}(T)$, is employed. $T$ is an integer variable that represents time. $s_{\text{in}}(T)$ randomly assumes '0' or '1' in every time step.

$$s_{\text{in}}(T) = 0 \text{ or } 1. \tag{1}$$

The middle-layer state is defined as a node vector $\mathbf{x}(T)$, consisting of $N$ elements.

$$\mathbf{x}(T) = \begin{pmatrix} x_1(T) \\ \vdots \\ x_N(T) \end{pmatrix}. \tag{2}$$

The time evolution of the node is determined from the input at the present time and the past state of the node.

$$\mathbf{x}(T+1) = f(\mathbf{x}(T), s_{\text{in}}(T+1)). \tag{3}$$

Here, $f$ is a function for time evolution. Then, the output $y_{\text{out}}(T)$, is defined as the inner product of the time-independent weight vector $\mathbf{W}_{\text{out}}$, and the node vector $\mathbf{x}(T)$.

$$y_{\text{out}}(T) = \sum_{i=1}^{N} W_i x_i(T) = \mathbf{W}_{\text{out}} \mathbf{x}(T) = \begin{pmatrix} W_1 & \cdots & W_N \end{pmatrix} \begin{pmatrix} x_1(T) \\ \vdots \\ x_N(T) \end{pmatrix}. \quad (4)$$

$$\mathbf{W}_{\text{out}} = \begin{pmatrix} W_1 & \cdots & W_N \end{pmatrix}. \quad (5)$$

The training data $y_{\text{train}}(T)$ is prepared to optimize the system. $\mathbf{W}_{\text{out}}$ is determined for $y_{\text{out}}(T)$, reproducing $y_{\text{train}}(T)$, and is selected to minimize the mean squared error (*MSE*) between $y_{\text{out}}(T)$ and $y_{\text{train}}(T)$. The *MSE* is expressed as follows:

$$MSE = \frac{1}{L} \sum_{T=1}^{L} \left( y_{\text{train}}(T) - y_{\text{out}}(T) \right)^2 = \frac{1}{L} \sum_{T=1}^{L} \left( y_{\text{train}}(T) - \mathbf{W}_{\text{out}} \mathbf{x}(T) \right)^2. \quad (6)$$

$\mathbf{W}_{\text{out}}$ is optimized using $L$ time steps. In this paper, $L$ is 2000. A pseudoinverse matrix $\mathbf{X}^{-1}$, is used for optimization.

$$\mathbf{W}_{\text{out}}^{\text{t}} = \mathbf{X}^{-1} \mathbf{y}_{\text{train}}. \quad (7)$$

The optimization of the output weight vector $\mathbf{W}_{\text{out}}$, is called *learning*. In this paper, a time-independent constant is added to $x_{N+1}(T)$ as a bias term, in addition to $x_1(T)$ to $x_N(T)$.

## B. Figures-of-merit for reservoir computing

In this paper, two types of task are employed for learning. One is a short-term memory (STM) task [28] for characterizing the memory effect in the system. The training data for the short-term memory task is expressed as follows:

$$y_{\text{train, STM}}(T, T_{\text{delay}}) = s_{\text{in}}(T - T_{\text{delay}}). \quad (8)$$

Here, $s_{\text{in}}(T)$ are random pulses, which are described later. It is feasible to obtain a finite memory effect, even if the system is completely linear. Therefore, we need another task to characterize the computing capability. In this paper, we employ the parity check (PC) task [28], in addition. The training data for the parity check task requests the parity of the input sum. Parity check is used for characterizing the type of non-linearity in the system and is expressed as follows:

$$y_{\text{train, PC}}(T, T_{\text{delay}}) = s_{\text{in}}(T - T_{\text{delay}}) + s_{\text{in}}(T - T_{\text{delay}} + 1) + \cdots + s_{\text{in}}(T) \, (\text{mod} \, 2). \tag{9}$$

After learning with the training data, the correlation between the output and training data is evaluated using the following equation:

$$Cor.(T_{\text{delay}})^2 = \frac{Cov.(y_{\text{train}}(T, T_{\text{delay}}), y_{\text{out}}(T))^2}{Var.(y_{\text{train}}(T, T_{\text{delay}})) Var.(y_{\text{out}}(T))}. \tag{10}$$

Here, *Cor.*, *Cov.*, and *Var.* are the correlation, covariance, and variance, respectively. In this paper, $Cor.^2$ is evaluated during 2000 time-steps, after learning with 500 time-steps. $Cor.^2$ assumes values from 0 to 1, with a larger $Cor.^2$ indicating better learning. Moreover, the capacity, *C,* is defined as the integration of $Cor.^2$, which can be used for evaluating the figures-of-merit for RC.

$$\text{Capacity}: C \equiv \sum_{T_{\text{delay}}=1}^{T_{\text{delay, max}}} Cor.(T_{\text{delay}})^2. \tag{11}$$

$T_{\text{delay, max}}$ should be sufficiently large. In our calculation, *Cor.* is always less than 0.01, when $T_{\text{delay}}$ is more than 10. Therefore, we set $T_{\text{delay, max}} = 30$. In this paper, we define $C_{\text{STM}}$ as the capacity for short-term memory and $C_{\text{PC}}$ as the capacity for parity check.

### C. Magnetic tunnel junction system

Figure 2 shows the schematics of the RC simulation using MTJs. An MTJ contains an insulating tunneling barrier layer with two ferromagnetic layers. For ferromagnetic layer-1 called the reference layer, the magnetization direction is designed to be fixed. This can be done by the exchange bias effect using antiferromagnetic materials, such as PtMn and IrMn [29], or magnetic anisotropy energy. In this study, the magnetization direction of layer-1 is fixed perpendicular to the film plane. For ferromagnetic layer-2 called the free layer, the magnetization direction is not fixed and can be controlled by the current [6,7]- or voltage [8]-driven spin-torque. The MTJ device resistance reflects the magnetization direction of **s**$_2$.

The spin-dynamics in ferromagnetic layer-2 follows the Landau-Lifshitz-Gilbert (LLG)

equation with spin-transfer torque [30], where a thermal fluctuation in ferromagnetic layers [31] is not included.

$$\frac{d\mathbf{s}_2}{dt} = -\gamma_0 \mathbf{s}_2 \times \mathbf{H}_{\text{eff}} - \alpha \mathbf{s}_2 \times \frac{d\mathbf{s}_2}{dt} + \frac{P}{1+P^2 \mathbf{s}_1 \cdot \mathbf{s}_2} \frac{I}{-e} \frac{\hbar}{2} \mathbf{s}_2 \times (\mathbf{s}_1 \times \mathbf{s}_2). \qquad (12)$$

Here $\mathbf{s}_1$ and $\mathbf{s}_2$ represent the unit spin-vectors for ferromagnetic layers-1 and -2, respectively. $\gamma_0$ (<0) is the gyro magnetic ratio. $\alpha$ is the Gilbert damping constant. $P$ is the spin polarization in vicinity of the Fermi level in the ferromagnetic layers. $I$ (=$V_{\text{in}}/R$) is the electric current, where $V_{\text{in}}$ and $R$ are the input voltage and device resistance of the MTJ respectively. $\mathbf{H}_{\text{eff}}$ is the effective magnetic field in $\mathbf{s}_2$.

$$\mathbf{H}_{\text{eff}} = -\frac{1}{\gamma_0} \nabla U. \qquad (13)$$

Here, $U$ is the magnetization energy for ferromagnetic layer-2, which includes the external magnetic field $\mathbf{H}_{\text{ext}}$ and the magnetic anisotropy tensor $\hat{\mathbf{H}}_{\text{ani}}$.

$$U = \mu_0 M_S A \left( \mathbf{H}_{\text{ext}} \cdot \mathbf{s}_2 + \frac{1}{2} \mathbf{s}_2^t \cdot \hat{\mathbf{H}}_{\text{ani}} \cdot \mathbf{s}_2 \right) = \frac{1}{2} \mu_0 M_S A \mathbf{s}_2^t \cdot \hat{\mathbf{H}}_{\text{ani}} \cdot \mathbf{s}_2 \ (\because \mathbf{H}_{\text{ext}} = \mathbf{0}), \qquad (14)$$

$$\hat{\mathbf{H}}_{\text{ani}} = \begin{pmatrix} 0 & 0 & 0 \\ 0 & 0 & 0 \\ 0 & 0 & H_{azz} \end{pmatrix}. \qquad (15)$$

Here $\mu_0$, $M_S$ and $A$ are is the magnetic permeability in a vacuum, saturation magnetization and volume of the ferromagnetic layer-2, respectively. In the simulation, the external magnetic field is not applied. We assume uniaxial anisotropy perpendicular to the film plane. Here $H_{azz} > 0$ ($H_{azz} < 0$) shows in-plane (perpendicular) magnetic anisotropy. The device resistance of the MTJ varies as a function of the relative angle between the spins in the free and pinned layers.

$$R = \frac{R_{\text{AP}} R_{\text{P}}}{(R_{\text{AP}} + R_{\text{P}}) + (R_{\text{AP}} - R_{\text{P}})(\mathbf{s}_1 \cdot \mathbf{s}_2)}. \qquad (16)$$

$R_{\text{AP}}$ and $R_{\text{P}}$ are the resistances, when $\mathbf{s}_1$ and $\mathbf{s}_2$ are parallel and antiparallel, respectively. The time evolution of the MTJ resistance is characterized by sequential calculation using the fourth Runge-Kutta method. For evaluating the short-term memory and parity-check capacities, an input pulse voltage, $V_{\text{in}}$, corresponding to the computational input, $s_{\text{in}}(T)$, was applied to the MTJs, as depicted in Fig. 3(a). Figures 2(a) and 2(b) display the schematics of circuits with single and multiple MTJs, respectively. In this paper, the physical parameters listed in Table I are employed. It almost follows our previous experimental research [32].

### D. Reference calculation with echo-state network

Additionally, an echo-state network [13] is introduced for comparison with the system using MTJs, where the following function is employed for Eq. (3):

$$\mathbf{x}(T) = \tanh\left(\mathbf{W}\mathbf{x}(T-1) + \mathbf{W}_{\text{in}} s_{\text{in}}(T)\right). \tag{17}$$

The tanh function is used for componentwise projection. $\mathbf{W}$ and $\mathbf{W}_{\text{in}}$ are matrices, whose components are time-independent random values from (−1) to 1. We normalize by dividing each component of $\mathbf{W}$ by the spectral radius, $r$, obtained by singular value decomposition [14].

## III. RESULTS & DISCUSSION

Figure 3(a) depicts an example of the MTJ device resistance, under an input voltage, $V_{\text{in}}$. We employed a pulse voltage with binary values of $V_0$ (= −44 mV) and $V_1$ (= +44 mV) as $V_{\text{in}}$. These binary values of $V_{\text{in}}$ correspond to 0 and 1 in $s_{\text{in}}(T)$, respectively, in the RC learning and evaluation processes (Eq. (1)). The pulse width (20 ns in Fig. 3(a), for instance) corresponds to the discrete unit time step $T$. Because the device resistance is scalar, the node dimension is only one. However, the number of nodes can be increased by employing virtual nodes [33,34]. As shown in the inset of Fig. 3(a), the virtual nodes $x_1$ to $x_N$, are defined; these virtual nodes are further defined as a node vector $\mathbf{x}(T')$.

Figure 3(b) depicts the DC bias voltage dependence of the static MTJ device resistance. Under a DC bias voltage, the MTJ device resistance was collected after the spin-dynamics were damped. Under a positive bias voltage, the spin-polarized current flows from the free layer $\mathbf{s}_2$, to the pinned layer $\mathbf{s}_1$. Then, the spin-transfer effect induces auto-oscillation [35, 36] in $\mathbf{s}_2$. The relative magnetization angle between $\mathbf{s}_1$ and $\mathbf{s}_2$ increases, and an antiparallel-like magnetization configuration is realized. Therefore, the device resistance increases, when a positive bias voltage is applied. Under a negative bias voltage, a parallel-like magnetization configuration is induced, and the device resistance decreases. For the input pulse voltage in Fig. 3(a), the binary values $V_0$ and $V_1$, are defined as voltages that render the device resistance constant. As shown in Fig. 3(b), $V_0$ and $V_1$ vary as a function of the uniaxial magnetic anisotropy $H_{azz}$.

### A. Figures-of-merit for reservoir computing using a single MTJ

In this section, we present the figures-of-merit for RC, using a single MTJ device. The uniaxial magnetic anisotropy of the free layer $s_2$, is fixed as $H_{azz} = 1000$ Oe. Here positive value of $H_{azz}$ shows magnetic cell in MTJ is in-plane magnetized. Figure 4 shows the simulated data for evaluating the short-term memory and parity-check capacities, for a single MTJ. In Fig. 4, the input-voltage pulse width is 20 ns and the number of virtual nodes, $N$, is 50. Figure 4 (a) shows the typical simulation results for an input $s_{in}(T)$, training data for the short-term memory task $y_{train, STM}(T)$ and trained output $y_{out}(T)$, as a function of the time step. Similarly, Fig. 4 (b) shows the input $s_{in}(T)$, training data for parity check task $y_{train, PC}(T)$, and trained output $y_{out}(T)$. Here training data for the short-term memory task $y_{train, STM}(T)$ and parity check task $y_{train, PC}(T)$ are defined using Eqs. (8) and (9) at $T_{delay} = 1$, respectively. The output is calculated using the simulated MTJ resistance (see Fig. 3) and $W_{out}$ using Eq. (4). $W_{out}$ is trivially calculate using the definitions given by Eqs. (5)-(7).

Figures 4 (c) and d) depict the correlations (Eq. (10)) between $y_{out}$ and the training data as a function of $T_{delay}$. We used $y_{train\ STM}$ as the training data for short-term memory and $y_{train\ PC}$ for the parity check. $C_{STM}$ and $C_{PC}$ are defined as the numerical integration of the correlation, and as the capacity using training data for the short-term memory and parity check, respectively.

Figure 5 shows the $C_{STM}$ and $C_{PC}$, respectively, as functions of the input-voltage pulse width (Figs. 5(a) and (b)) and the number of virtual nodes $N$ (Fig. 5(c) and (d)). From Figs. 5(a) and (b), both $C_{STM}$ and $C_{PC}$ increase as the pulse width increases. When the pulse width is greater than 20 ns, $C_{STM}$ and $C_{PC}$ are nearly constant because, when the pulse width is less than 20 ns, the change in the magnetization direction is very small and the spin-dynamics cannot work as a reservoir. In Figs. 5(c) and (d), the dependence of $C_{STM}$ and $C_{PC}$, respectively, on the number of the virtual nodes $N$, are displayed, when the pulse width is fixed at 20 ns. From Figs. 5(c) and 5(d), we find that it is better to set the number of virtual nodes greater than 20.

### B. Figures-of-merit for reservoir computing using multiple MTJs

When multiple MTJs are employed for RC, higher figures-of-merit can be obtained. A schematic of a multiple MTJ circuit for RC is depicted in Fig. 2(b). Multiple MTJs are placed in parallel, and an identical pulse voltage is applied to all the MTJs. To construct nodes for RC, spatial multiplexing [37] is employed. The node vector $\mathbf{x}(T)$, is defined as a vector with $M \times N$ elements, where $M$ is the number of MTJs and $N$ is the number of virtual nodes in an MTJ.

$$\mathbf{x}_1(T) = \begin{pmatrix} x_{11}(T) \\ \vdots \\ x_{1N}(T) \end{pmatrix}, \mathbf{x}_2(T) = \begin{pmatrix} x_{21}(T) \\ \vdots \\ x_{2N}(T) \end{pmatrix}, \cdots, \mathbf{x}_M(T) = \begin{pmatrix} x_{M1}(T) \\ \vdots \\ x_{MN}(T) \end{pmatrix}$$
$$\rightarrow \mathbf{x}(T) \equiv \begin{pmatrix} x_{11}(T) & \cdots & x_{1N}(T) & x_{21}(T) & \cdots & x_{2N}(T) & \cdots & x_{M1}(T) & \cdots & x_{MN}(T) \end{pmatrix}^t \tag{18}$$

Figure 6 shows the $C_{STM}$ and $C_{PC}$ with multiple MTJs. The uniaxial anisotropy $H_{azz}$, of ferromagnetic layer-2 in each MTJ is listed in Table II. For instance, when four MTJs and $H_{azz, k}/H_{azz, k+1} = 2$ are employed, the uniaxial anisotropies of the MTJs are 1000 Oe, 500 Oe, 250 Oe, and 125 Oe, respectively, as shown in Fig. 6(a). Such variations in the anisotropies can be obtained by voltage-controlled magnetic anisotropy in the MTJs [38]. In this study, thermal fluctuation in ferromagnetic layer-2 is not included. For instance, thermal fluctuation energy at room temperature (26 meV) is negligibly small when compared to the magnetization energy from Eq. (14) (~10 eV) when the magnetic anisotropy energy is $H_{azz} = 1000$ Oe. Therefore thermal fluctuation can be comparable or less than the magnetization energy of ferromagnetic layer-2 at $H_{azz} < 3$ Oe. In such region, simulations assuming the ground state are not very correct, and a random magnetic field to reproduce the thermal fluctuation [31] should be included in the simulation.

Similar to a single MTJ, the binary values $V_0$ and $V_1$, for the input voltage are determined as shown in Fig. 3(b). Note that $V_0$ and $V_1$ vary as a function of the uniaxial anisotropy field, and the smallest absolute values of the saturation voltages are employed as $V_0$ and $V_1$ for RC with multiple MTJs; i.e., $V_0$ and $V_1$ are determined for the MTJ with the smallest uniaxial magnetic anisotropy field. Figures 6(b) and (c) display the $C_{STM}$ and $C_{PC}$, respectively, as functions of the anisotropy ratio $H_{azz, k}/H_{azz, k+1}$. In the simulation, the input-voltage pulse width is 20 ns and the number of virtual nodes for each MTJ is 50, for all MTJs. The maximum value of $C_{STM}$ increases as the number of MTJs ($M$) increase. Because each MTJ has a different uniaxial magnetic anisotropy field $H_{azz}$, it has a different response speed to external voltage/current. This variation in the response speed increases the $C_{STM}$ of the system. On the other hand, the increase in $C_{PC}$ is insignificant compared to that of the $C_{STM}$. Note that as there is no electric and/or magnetic interaction between the free layers of the MTJs, the $C_{PC}$ is insignificant. In Fig. 6(c), when $H_{azz, k}/H_{azz, k+1}$ is large, the $C_{PC}$ using multiple MTJs is less than that using a single MTJ. This is because the input-voltage pulse width of 20 ns is the best condition only for the parameters of a single MTJ ($H_{azz}$ = 1000 Oe, $V_1$ = 44 mV, $V_0$ = −44 mV).

Figure 7 shows the $C_{STM}$ and $C_{PC}$, under various condition. In Fig. 7, $H_{azz, k} / H_{azz, k+1}$ is fixed to 1.6. This is the best condition for the $C_{STM}$ with $M$ = 7. From Fig. 7(a), the $C_{STM}$ is maximum, around a pulse width of 20 ns. When the pulse width is lesser than 20 ns, the change in the magnetization direction by the spin-transfer torque is too small for performing as a reservoir. When the pulse width is greater than 20 ns, the spin-dynamics are almost damped during a unit time step, and such a condition is not preferable for RC. From Fig. 7(b), the best conditions for the $C_{STM}$ and $C_{PC}$ are not identical. This is because a relatively long pulse is required to induce non-linearity in the spin-dynamics, in multiple MTJs. Figures 7(c) and (d) depict the $N$ dependences of the $C_{STM}$ and $C_{PC}$, respectively, when $M = 7$, $H_{azz, k} / H_{azz, k+1} = 1.6$, and the pulse width = 20 ns. When $N$ is greater than four, both the $C_{STM}$ and $C_{PC}$ are nearly constant.

## C. Comparison with the echo-state network

The $C_{STM}$ and $C_{PC}$, using a multiple MTJ-system, are summarized in Fig. 8 (a); the pulse width = 20 ns and the virtual node number $N = 50$, for each MTJ. The data points from top to bottom are the data, when $H_{azz, k}/H_{azz, k+1}$ changes from 1.1 to 3.0. The result of the echo-state network, using the tanh function, is shown in Fig. 8 (b). The data points from top to bottom are the data, when the spectrum radius, $r$, of **W** (see Eq. (17)) varies from 0.05 to 2.0. From Fig. 8, it can be observed that a high performance can be obtained for RC using 5-7 MTJs, similar to an echo-state network with 20-30 nodes. In terms of the total number of virtual nodes in the system ($M \times N$), 35 nodes ($7 \times 5$) of an MTJ-system are comparable to 20-30 nodes of an echo-state network (see also Figs. 7(c) and (d)). Although the $C_{PC}$ increases slightly as $M$ increases, we can obtain a large $C_{PC}$, if there are magnetic and/or electrical interactions between the free layers in each MTJ.

## IV. CONCLUSION

In this research, we demonstrated RC, using the spin-dynamics in MTJs. With RC using 5-7 MTJs, we can obtain a high performance similar to that of an echo-state network using tanh functions with 20-30 nodes. If there are magnetic and/or electrical interactions between the free layers in each MTJ, higher performance can be obtained.

## ACKNOWLEDGEMENTS

We thank E. Tamura, K. Shimose, S. Hasebe, and M. Goto of Osaka University, T. Taniguchi of AIST for the discussions. Part of this work was supported by JSPS KAKENHI (Nos. JP18H03880, JP26103002, JP16H02211, JP16KT0019, JP15K16076, and JP26880010) and the Ministry of Internal

Affairs and Communications. This work was also supported by JST-ERATO (JPMJER1601), JST-CREST (JPMJCR1673), and JST-PRESTO (JPMJPR1668 and JPMJPR15E7), Japan.

Table I. Physical parameter set of ferromagnetic layer-2 for RC with a single MTJ (Fig. 2(a)).

| Parameter | Value |
| --- | --- |
| Gilbert damping constant (layer-2): $\alpha$ | 0.009 |
| Uniaxial anisotropy (layer-2): $H_{azz}$ | 1000 Oe |
| Saturation magnetization (layer-2): $M_S$ | 1375 emu/cc |
| Volume (layer-2): $A$ | 23500 nm$^3$ ($\phi$122 nm × 2 nm) |
| Resistance in parallel: $R_P$ | 210 Ω |
| Resistance in antiparallel: $R_{AP}$ | 390 Ω |

Table II. Variation for uniaxial magnetic anisotropy for RC with multiple MTJs (Fig. 2(b)).

| $H_{azz, k} / H_{azz, k+1}$ | $H_{azz, 1}$ | $H_{azz, 2}$ | $H_{azz, 3}$ | … | $H_{azz, 7}$ |
|---|---|---|---|---|---|
| 1.0 | 1000 Oe | 1000 Oe | 1000 Oe | … | 1000 Oe |
| 1.1 | 1000 Oe | 909.1 Oe | 826.4 Oe | … | 564.5 Oe |
| 1.2 | 1000 Oe | 833.3 Oe | 694.4 Oe | … | 334.9 Oe |
| … | … | … | … | … | … |
| 2.9 | 1000 Oe | 344.8 Oe | 118.9 Oe | … | 1.7 Oe |
| 3.0 | 1000 Oe | 333.3 Oe | 111.1 Oe | … | 1.4 Oe |

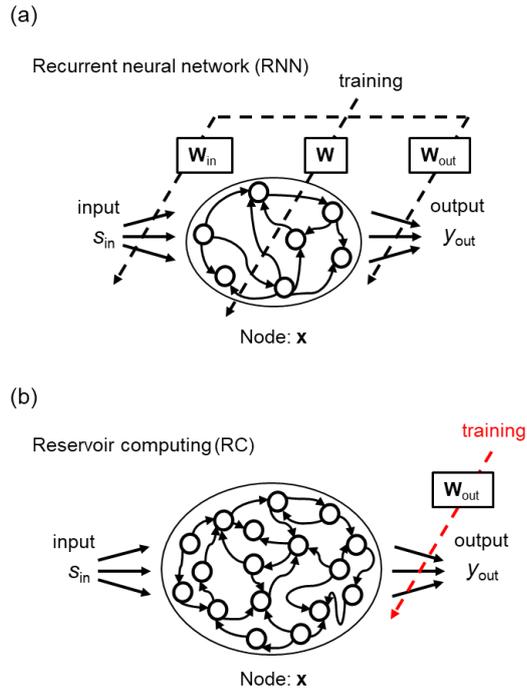

FIG. 1

FIG. 1. Concept of (a) recurrent neural network (RNN) and (b) reservoir computing (RC).

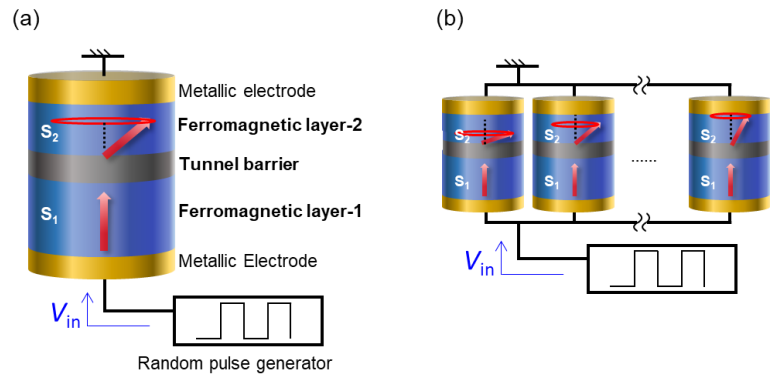

FIG. 2

FIG. 2. (a) Schematic of a RC system using the spin-dynamics in a magnetic tunnel junction (MTJ) and (b) system with multiple MTJs. In the MTJs, the spin direction of the ferromagnetic layer-2 ($\mathbf{s}_2$) can be controlled by the input bias voltage $V_{in}$, whereas that of the ferromagnetic layer-1 ($\mathbf{s}_1$) is fixed.

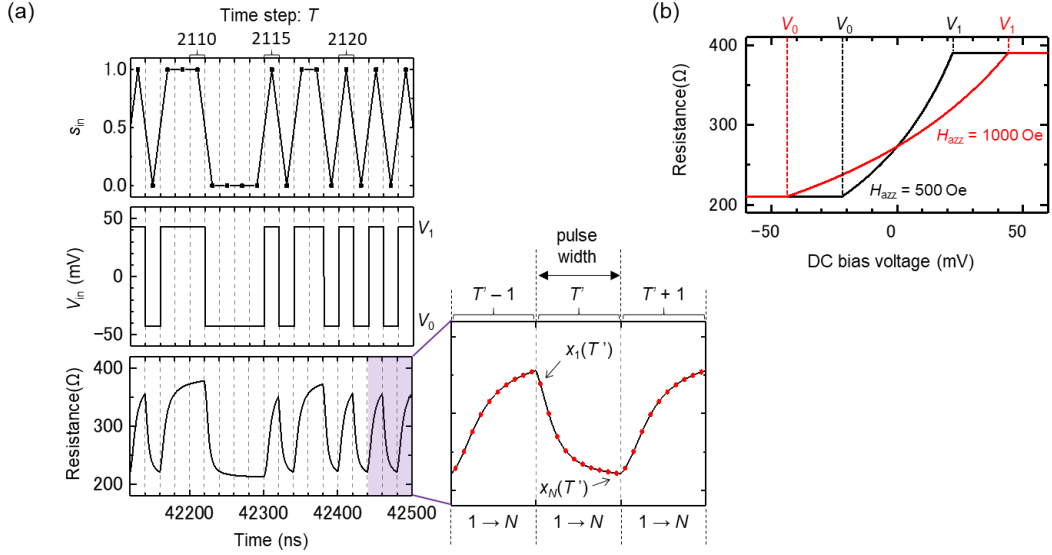

FIG. 3.

(a) Input $s_{in}(T)$, input bias voltage to the MTJ device $V_{in}$, and MTJ device resistance as a function of time. Typical characteristics during the learning and evaluation processes. We define the virtual nodes ($x_1(T)$ to $x_N(T)$) as shown in the inset and the (b) MTJ device resistance as a function of the static input DC bias voltage. The black and red plots indicate the resistances, when the values of the uniaxial anisotropy fields are 500 Oe and 1000 Oe, respectively. $V_0$ and $V_1$ are voltages that render the device resistance constant.

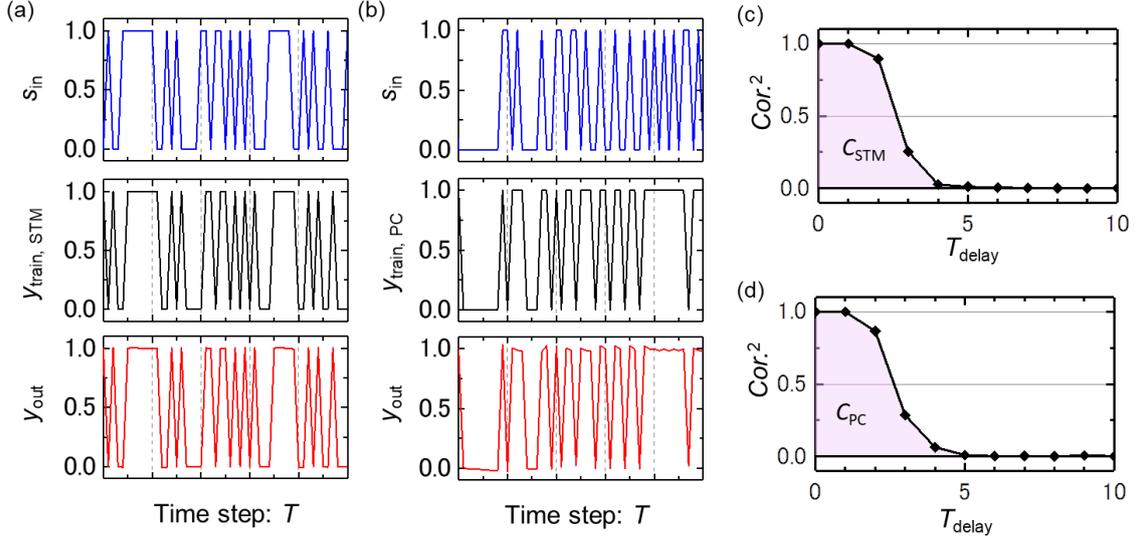

FIG. 4

FIG. 4. (a) Input $s_{in}$, output for training $y_{train\ STM}$ (Eq. (8) with $T_{delay} = 1$), and trained output $y_{out}$, for evaluating the short-term memory task, (b) Input $s_{in}$, output for training $y_{train\ PC}$, (Eq. (9) with $T_{delay} = 1$), and trained output $y_{out}$, for evaluating the parity check task, (c) Correlation using Eq. (10); the integrated values are defined as the short-term memory capacity ($C_{STM}$), (d) Correlation using Eq. (10); the integrated values are defined as the parity-check capacity ($C_{PC}$). The input-voltage pulse width = 20 ns and the number of virtual nodes $N = 50$.

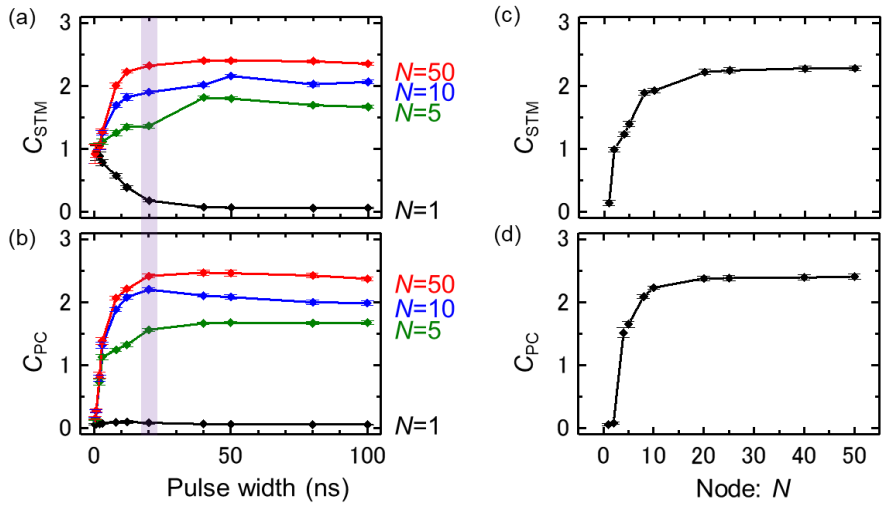

FIG. 5

FIG. 5. Results of RC using single MTJ. (a) Short-term memory capacity ($C_{\text{STM}}$) as a function of the input-voltage pulse width, (b) Parity-check capacity ($C_{\text{PC}}$) as a function of the input-voltage pulse width; $N$ is number of virtual nodes in the MTJ, (c) $C_{\text{STM}}$, and (d) $C_{\text{PC}}$ as functions of the virtual-node number, where the input-voltage pulse width is fixed to 20 ns.

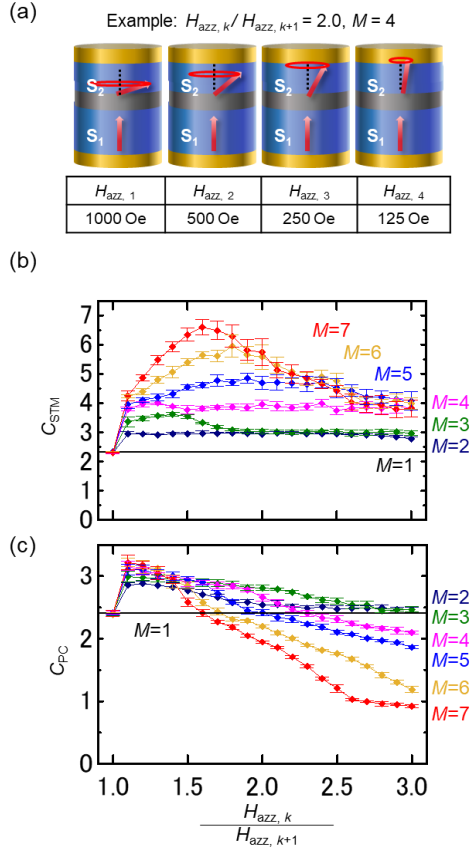

FIG. 6

FIG. 6. Results of RC using multiple MTJs. (a) Example of a parameter set for multiple MTJs, when the ratio of the uniaxial magnetic anisotropy in each MTJ ($H_{azz, k+1} / H_{azz, k}$) = 2 and the number of MTJs ($M$) = 4, (b) Short-term memory capacity ($C_{STM}$), and (c) parity check capacity ($C_{PC}$) as functions of $H_{azz, k+1} / H_{azz, k}$. The input-voltage pulse width = 20 ns and the virtual nodes for each MTJ ($N$) = 50.

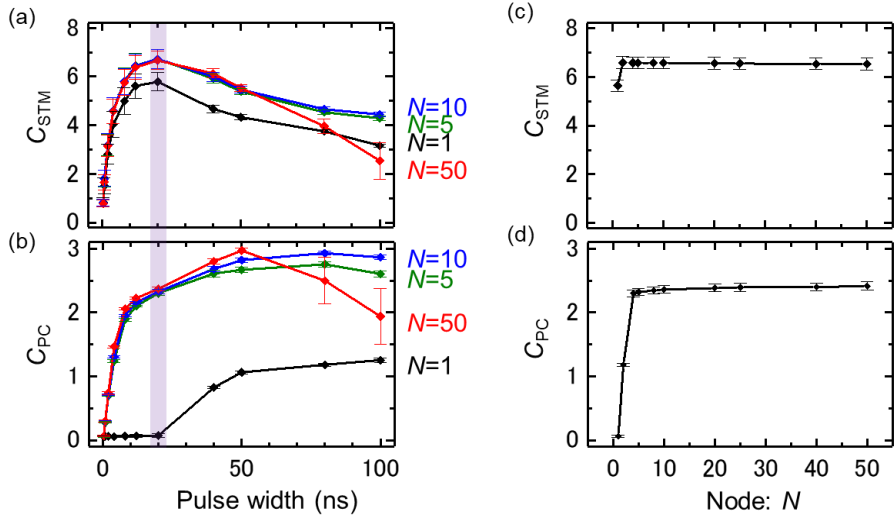

FIG. 7

FIG. 7. Results of RC using multiple MTJs. (a) Short-term memory capacity ($C_{STM}$), and (b) Parity check capacity ($C_{PC}$) with seven MTJs ($M=7$) as functions of the input-voltage pulse width, (c) $C_{STM}$, and (d) $C_{PC}$ as functions of the virtual node number of each MTJ ($N$), under an input-voltage pulse width of 20 ns. The uniaxial magnetic anisotropy ratio ($H_{azz,\,k+1} / H_{azz,\,k}$) = 1.6.

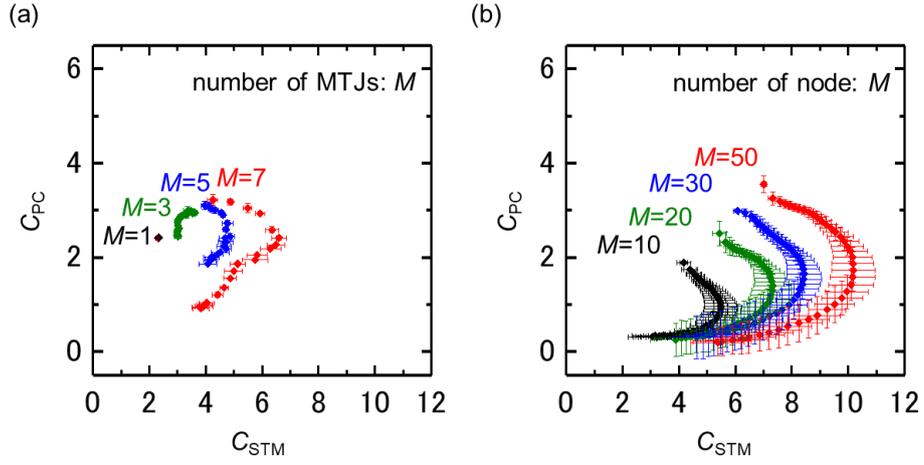

FIG. 8

FIG. 8. Plots showing $C_{STM}$ vs $C_{PC}$ in the (a) MTJ-system and (b) echo-state network with Eq. (17). In the MTJ-system, the pulse width and virtual nodes of each MTJ ($N$) are fixed to 20 ns and 50, respectively.